\documentclass[12pt]{article}
\usepackage{amssymb}
\usepackage{graphicx}

\usepackage{amssymb}
\usepackage{mathdots}
\usepackage{setspace}
\usepackage{stmaryrd}
\usepackage{amsmath}

\DeclareMathAlphabet\mathbit
    \encodingdefault\rmdefault\bfdefault\itdefault
\DeclareOldFontCommand{\bi}{\normalfont\bfseries\itshape}{\mathbit}

\newcommand{\be}{\begin{equation}}
\newcommand{\ee}{\end{equation}}

\def\fakebold#1{\relax\ifvmode\leavevmode\fi%
\ifmmode%
\setbox0=\hbox{$#1$}%
\else%
\setbox0=\hbox{#1}%
\fi%
\kern-.02em\copy0 \kern-\wd0%
\kern .04em\copy0 \kern-\wd0%
\kern-.0125em\raise.02em\box0%
}%

\begin{document}

\title{SHEAR INSTABILITY IN SKIN TISSUE}

\author {Pasquale CIARLETTA, Michel DESTRADE, Artur L. GOWER} 

\date{2013}

\maketitle

\begin{abstract}
We propose two toy-models to describe, predict, and interpret the
wrinkles appearing on the surface of skin when it is sheared. With
the first model, we account for the lines of greatest tension
present in human skin by subjecting a layer of soft tissue to a
pre-stretch, and for the epidermis by endowing one of the layer's
faces with a surface tension. For the second model, we consider an
anisotropic model for the skin, to reflect the presence of stiff
collagen fibres in a softer elastic matrix. In both cases, we find
an explicit bifurcation criterion, linking geometrical and material
parameters to a critical shear deformation accompanied by small
static wrinkles, with decaying amplitudes normal to the free surface
of skin.

\end{abstract}


\section{Introduction}

When the skin is pinched, wrinkles appear quite early on its
surface. The same phenomenon occurs when the skin is sheared, i.e.
pinched with one finger moving in one direction and the other fixed
or moving in the opposite direction. In fact, pinching is one of the
tests performed by dermatologists and surgeons \cite{Wald02} when
trying to assess the direction of greatest tension \cite{Krai51} in
the neighborhood of a site of interest. Sometimes called \emph{lines of cleavage} \cite{Cox41}, the orientations of the lines of
greatest tension are crucial to the way a scar heals. For a cut
across the lines, the lips of a wound will be pulled away from one
another during the healing process, while they will be drawn
together if the cut has occurred parallel to the lines. In one case
the resulting scar can be quite unsightly, in the other it is almost
invisible. In this paper we investigate the mechanical stability of
two toy models for the human skin under shear in its plane, and view
the onset of small-amplitude, unstable solutions as a prototype for
skin wrinkling.

\begin{figure}[!ht]
\centerline{\includegraphics[height=5cm]{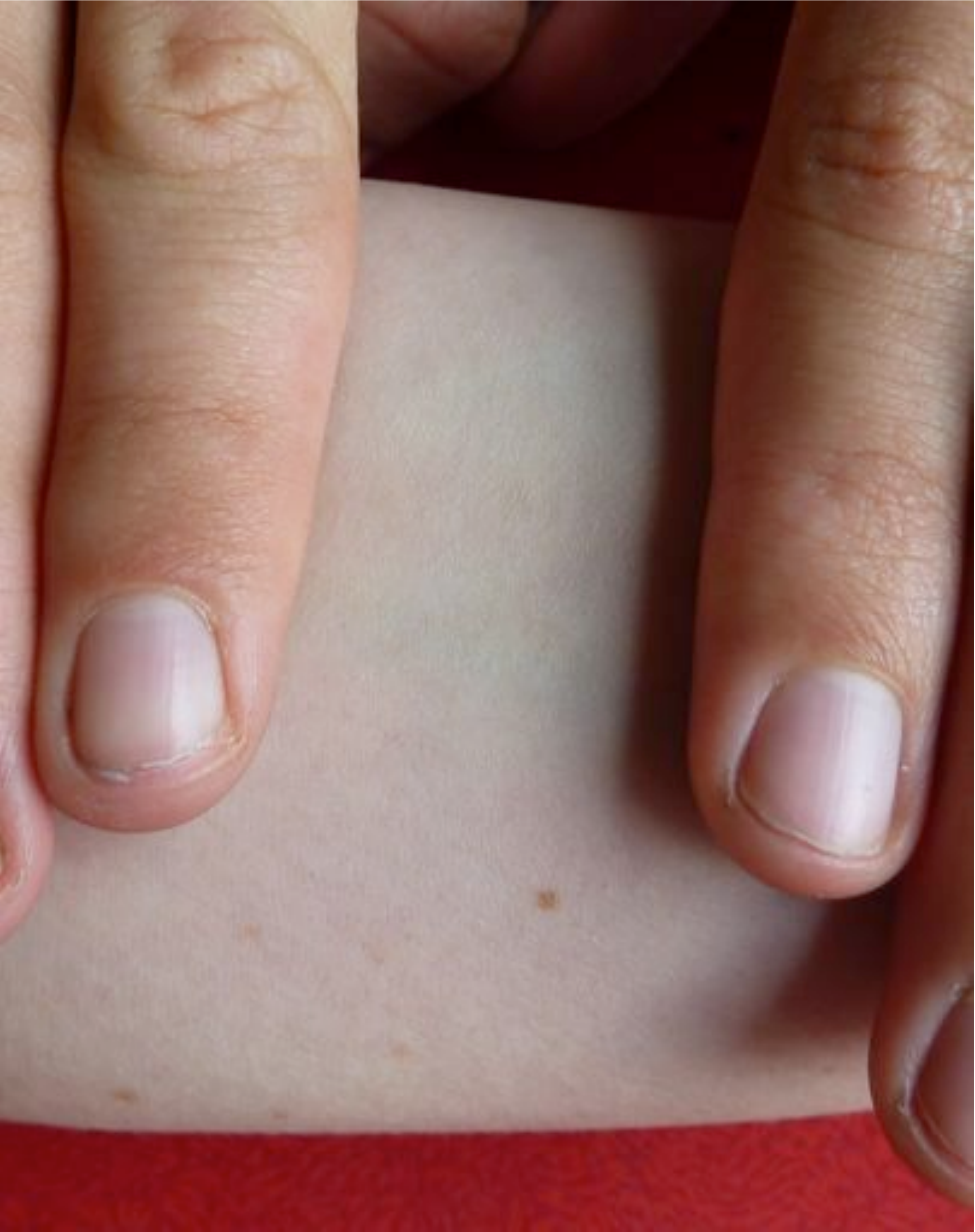} \qquad\includegraphics[height=5cm]{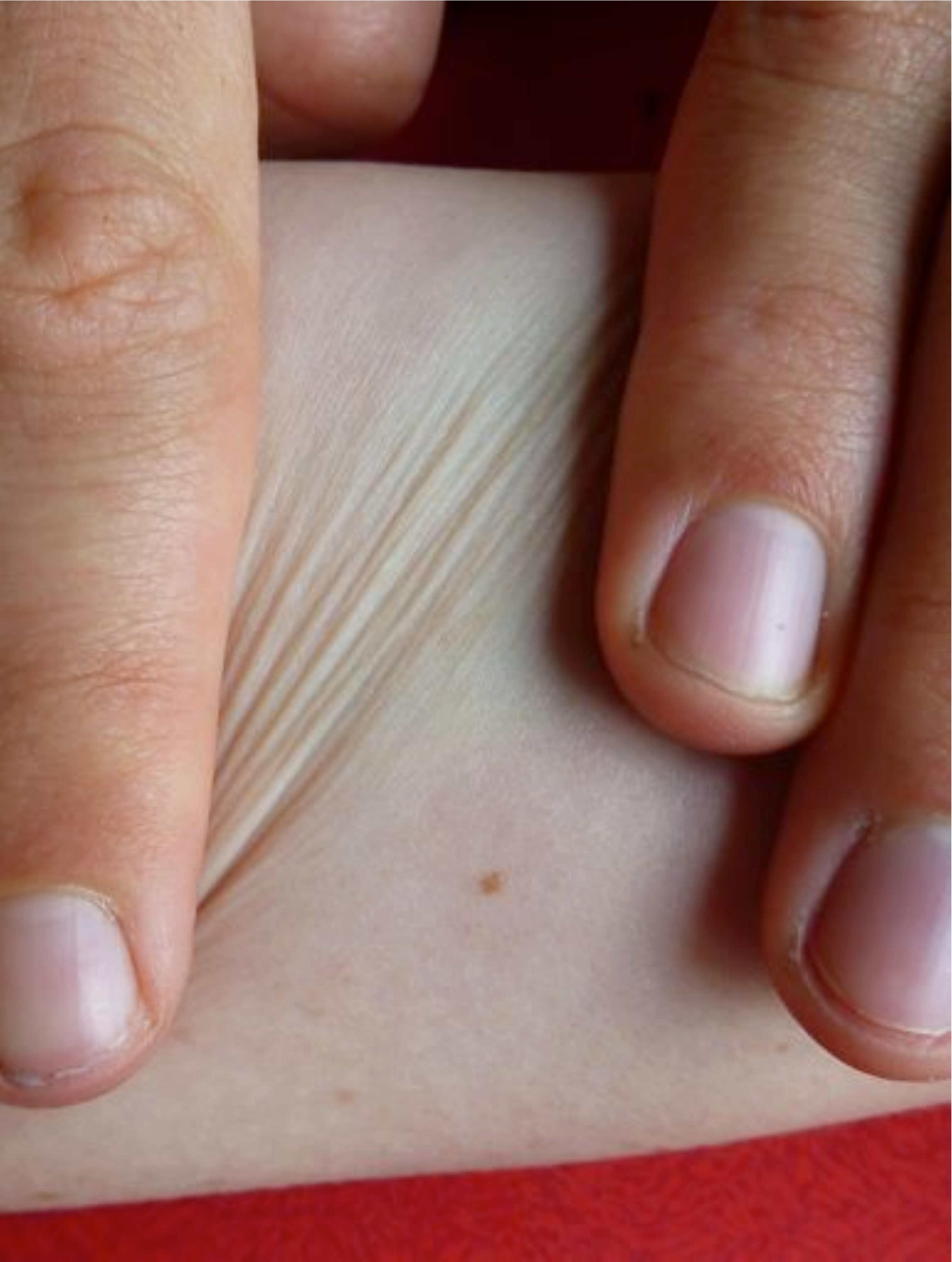}}
 \caption{{\small Shearing the forearm skin across the lines of cleavage (which run along the length of the arm) results in early onset of small-amplitude wrinkles.}}
\label{skin1}
\end{figure}

Of course, skin is a complex, multi-faced organ, and it is not
easily, nor perhaps realistically, modeled. In Section
\ref{Section2}, we first view it as an initially isotropic, neo-Hookean
layer of finite thickness. {Although a two-layered model of 
skin would be more realistic, it would greatly complicate the theoretical analysis of the shear instability properties. 
Therefore, we prefer to consider an epidermis of vanishing 
thickness on top of a hyperelastic dermis, by defining a proper surface 
elastic energy.} In other words we let one of the
layer's faces be a material curve endowed with intrinsic elastic
properties associated with extensibility, but no bending stiffness
(see \cite{StOg97} for a rigorous exposition of such elastic
coatings). We account for the lines of greatest tension by imposing
a finite plane pre-stretch in a given direction; in other words, we
simulate those lines through \emph{strain-induced anisotropy}. Then
we investigate whether surface energy and pre-stretch promote or
attenuate the appearance of wrinkles when the layer is subject to
simple shear in the direction of the cleavage lines.

In Section \ref{Shear instability of a fibre-reinforced skin
tissue}, we then view skin as being intrinsically \emph{anisotropic}
that is, we switch to the point of view that lines of greatest
tension are due to the presence of families of parallel bundles of
stiff collagen fibres imbedded in a softer elastin matrix. The
introduction of even the simplest anisotropy -- transverse isotropy
due to a single privileged direction -- complicates the equations of
incremental instability greatly, and we thus restrict attention to a
homogeneous solid without surface tension. We also omit finite-size
effects by considering a half-space instead of a layer, and thus by
focusing on the Biot surface instability phenomenon \cite{Biot63}.
Here, the anisotropic contribution to the stored energy is that
recently proposed by Ciarletta et al.\cite{ciarlettaJMBBM}. Its
polyconvexity ensures good properties from the physical point of
view, such as strong ellipticity in compression, in contrast to the
standard reinforcing model used recently by Destrade et
al.\cite{destrade08} for the same stability study.


\section{Shear instability for a neo-Hookean layer with surface coating}
\label{Section2}


First we consider an isotropic elastic material of finite  thickness
$H$, undergoing an homogeneous shear. In order to mimic the response
of human skin, we incorporate the presence of a residual stretch
$\lambda_\text{res}$ along the main  cleavage lines, so  that the
base deformation field reads
\begin{equation}
x= \lambda_\text{res} X + (K/\lambda_\text{res}) Y, \qquad y=
Y/\lambda_\text{res}, \qquad z=Z, \label{def}
\end{equation}
where $\mathbf x$ is the current position of a material point which
was at $\mathbf X$ in the reference configuration, and $K$ is a
constant. Hence, we see that the deformation can be decomposed as a
plane stretch of amount $\lambda_\text{res}$ followed by a simple
shear of amount $K$, with deformation gradient $\mathbf F$ written
as
\begin{equation}
\label{F} \mathbf F = \begin{bmatrix} \lambda_\text{res} &
K/\lambda_\text{res} & 0 \\ 0 & 1/\lambda_\text{res} & 0 \\ 0 & 0 &
1 \end{bmatrix}
 =  \begin{bmatrix} 1 & K & 0 \\ 0 & 1 & 0 \\ 0 & 0 & 1 \end{bmatrix}
  \begin{bmatrix} \lambda_\text{res} & 0 & 0 \\ 0 & 1/\lambda_\text{res} & 0 \\ 0 & 0 & 1 \end{bmatrix}.
  \end{equation}

  We note that the layer's thickness $H$ remains unchanged through the deformation.
For simplicity, we take the layer to be made of an isotropic
neo-Hookean incompressible material with a surface energy at the
free boundary $ z=0$, so that its total strain energy $W$ reads
\begin{equation}
W = \frac{\mu}{2} \iiint \left({\rm tr} \ {\bf b} -3 \right) \text
dX \ \text dY \ \text dZ + \iint \left(\gamma |{\bf x}_{,X} \times{\bf
x}_{,Y}|+\frac{\mu_s}{2}(|{\bf x}_{,X}|^2+|{\bf
x}_{,Y}|^2-2) \right) \text dX \ \text dY, \label{const}
\end{equation}
where $\mu$ is the shear modulus, $\mathbf b = \mathbf{FF}^T$ is the
left Cauchy-Green deformation tensor, {$\gamma$ is the surface tension coefficient, $\mu_s$ is the elastic shear modulus per surface unit}, and the comma denotes partial derivative. {This is akin to
endowing one of the material boundary of the layer with a surface
energy with a term proportional to changes in area (as is often done in
fluid mechanics, see e.g. \cite{Laut11}), and another contribution depending on the elastic deformation of the surface. 
This last term is proportional to a stretch measure of the surface deformation tensor, 
which is chosen for invariance requirements because we consider this elastic 
layer as  a hemitropic film \cite{Ogden99}. }

Now from the constitutive assumptions in
Eq.(\ref{const}), we find that $\boldsymbol \sigma$, the Cauchy
stress tensor corresponding to the large deformation in
Eq.\eqref{def}  is given by
\begin{equation}
\boldsymbol \sigma= \mu {\bf b} - p{\bf I},
\end{equation}
where $p$ is a Lagrange multiplier due to the constraint of incompressibility. 
Writing that  the boundary at $z=0$ is traction-free fixes the value of $p$ as $p=\mu$.

Straightforward calculations reveal that principal stretches of the
deformation field in Eq.(\ref{def}) are $\lambda_k$ ($k=1,2,3$)
given by \cite{DeOg05}
\begin{equation} \label{l1}
\lambda_{1} \pm \lambda_2  = \sqrt{(\lambda_\text{res} \pm
\lambda_\text{res}^{-1})^2 + K^2\lambda_\text{res}^{-2}},
\qquad\lambda_3=1,
\end{equation}
and that the Eulerian principal axes $(x_1, x_2)$ are obtained after
an anti-clockwise rotation of angle $\phi$ of the in-plane
coordinate axes about the $z$ axis, where
\begin{equation}
\tan(2 \phi)=
\frac{2K}{\lambda_\text{res}^2-\lambda_\text{res}^{-2}(1-K^2)}.
\label{vart}
\end{equation}
From Eq.\eqref{l1}, we note that
\begin{equation} \label{xi}
\lambda_2=\lambda_1^{-1}, \qquad K =
\lambda_\text{res}\sqrt{(\lambda_{1}-\lambda_{1}^{-1})^2
-(\lambda_\text{res}-\lambda_\text{res}^{-1})^2}.
\end{equation}

We now look for a perturbation solution in the neighbourhood of the
large deformation \eqref{def}, using the theory of incremental
deformations \cite{Ogde97}. Hence we call $\mathbf u = \mathbf
u(x_1, x_2, x_3)$ the incremental displacement field, for which the
incremental incompressibility condition imposes that
\begin{equation}
u_{i,i}=0. \label{inc1}
\end{equation}
The  constitutive equation for the components of the incremental
nominal stress $\mathbf{\dot S}$ reads in general as \cite{Ogde97},
\begin{equation}
\label{S1} \dot S_{ji} = L_{jikl} \, u_{k,l} + p\, u_{j,i} - \dot p
\, \delta_{ji},
\end{equation}
where $\dot p$ is the increment in the Lagrange multiplier, and $\textit{L}$ is the
fourth-order tensor of instantaneous moduli, i.e. the push-forward
of the fixed reference elasticity tensor.  In the absence of body
forces, we can therefore write the equilibrium equation of the
incremental nominal stress ${\bf{\dot S}}$ as
\begin{equation}
\label{divS} (\text{div} \, {\bf{\dot S}})_i= \dot{ S}_{ji,j}=
\mathrm 0.
\end{equation}
In the case of a neo-Hookean material, it is easy to check that the
components of $L$ are simply $L_{jikl}=\mu \delta_{jk} b_{il}$ so
that Eq.(\ref{divS}) in the coordinate system aligned with the
Eulerian principal axes takes the following simplified form
\begin{equation}
-\dot p,_1+\mu \lambda_1^2 u_{1,ii}=0, \qquad -\dot p,_2+\mu
\lambda_2^2 u_{2,ii}=0, \qquad -\dot p,_3+\mu \lambda_3^2
u_{3,ii}=0, \label{pi}
\end{equation}
Differentiating these incremental equilibrium equations with respect
to $x_1$, $x_2$, and $x_3$, respectively, and using the incremental
incompressibility condition in Eq.(\ref{inc1}), we find that
\begin{equation} \label{plapl}
\nabla^2 \dot p=0,
\end{equation}
that is, the incremental Lagrange multiplier is a Laplacian field
\cite{Flav63}.

Now, for the incremental \emph{boundary conditions}, we consider
that the bottom  $z=x_3=-H$ of the layer is fixed (clamped
condition):
\begin{equation}
u_i(x_1,x_2,-H)=0, \label{bcH}
\end{equation}
while the top face $z=x_3=0$ remains free of incremental traction:
\begin{align}
& u_{1,3}+u_{3,1}-\mu_s \lambda_1^2u_{1,11}-\mu_s \lambda_1^{-2}u_{1,22}=0,
\notag \\
& u_{2,3}+u_{3,2}-\mu_s \lambda_1^2u_{2,11}-\mu_s \lambda_1^{-2}u_{2,22}=0,
\notag \\
& - \dot p + 2\mu u_{3,3} - \gamma\left(u_{3,11}
+u_{3,22}\right) -\mu_s \lambda_1^2u_{3,11}-\mu_s \lambda_1^{-2}u_{3,22}=0. 
\label{bc3}
\end{align}

We search for solutions to Eqs.(\ref{inc1},\ref{pi},\ref{plapl}) in
the following form:
\begin{equation}
\left\{u_1,  u_2, u_3, \dot p\right\}=\left\{U_1(x_3), U_2(x_3),
U_3(x_3), ik P(x_3)\right\}  e^{ik(\cos\theta \, x_1 + \sin \theta
\, x_2)}, \label{sep}
\end{equation}
corresponding to the occurrence of plane wrinkles with wavenumber
$k$, forming an angle $\theta$ with the direction of maximum
extension. It is easy to show that a solution in the form of
Eq.(\ref{sep}) is given by \cite{Flav63}
\begin{align}
& U_1(x_3)= \cos\theta\left( a_1 e^{-k x_3}+a_2 e^{-qk x_3}+a_3 e^{k
x_3}+a_4 e^{qk x_3}\right), \notag
\\
& U_2(x_3)= \sin\theta\left( a_1 e^{-k x_3}+a_2 e^{-qk x_3}+a_3 e^{k
x_3}+a_4 e^{qk x_3}\right), \notag
\\
& U_3(x_3)= i\left( a_1 e^{-k x_3}+{a_2} e^{-qk x_3}/q - a_3 e^{k
x_3} - {a_4} e^{qk x_3}/q\right), \notag
\\
& P(x_3)= - \mu(1-q^2)\left( a_1 e^{-k x_3}+a_3 e^{k x_3}\right),
\label{P}
\end{align}
where  $a_1,a_2,a_3,a_4$ are yet arbitrary constants, and $q$ is
fixed by imposing Eqs.(\ref{pi}), as
\begin{equation}
q=  \sqrt{\lambda_1^2 \cos^2 \theta +\lambda_1^{-2} \sin^2 \theta}.
\end{equation}
Using Eqs.\eqref{P}, it can be checked that only four independent
boundary conditions result from Eqs.\eqref{bc3}. Setting ${\bf a} =
[a_1,a_2,a_3,a_4]^T$, they can be written in the following matrix
form:
\begin{equation}
{\bf Q} {\bf a}= \mathbf 0,
\end{equation}
where the components of the matrix  ${\bf Q}$  are
\begin{equation}
{\small
 \begin{bmatrix}
2q -kq^2L_\text{el} & 1+q^2  -kq^2L_\text{el} & -2q-kq^2L_\text{el} &-1-q^2-kq^2L_\text{el} \\
q(-1-q^2 +kL_\text{cap} +kq^2L_\text{el} ) & -2q+kL_\text{cap}+kq^2L_\text{el}  & q(-1-q^2
-kL_\text{cap} -kq^2L_\text{el} ) &
-2q-kL_\text{cap}-kq^2L_\text{el} \\
e^{kH} & e^{qkH} & e^{-kH} & e^{-qkH}\\
e^{kH} & e^{qkH}/q & -e^{-kH} & -e^{-qkH}/q
\end{bmatrix}.
}
\end{equation}
Here   $L_\text{cap}:=\gamma/\mu$, $L_\text{el}:=\mu_s/\mu$ are the \emph{characteristic
capillary \textbf{and elastic} lengths} of the material, respectively. The resulting condition for the
wrinkling instability is  $\det {\bf Q}=0$. After lengthy
manipulations not reproduced here, it is possible to show that the
earliest onset of instability occurs at $\theta=0$ that is, when the
wrinkles are aligned with the direction where the greatest stretch,
$\lambda_2$, takes place. There, $q=\lambda_1$, and the \emph{dispersion
relation} reduces to
\begin{multline}
k\lambda_1^2 L_\text{el}\left\{\left[2kL_\text{el}\lambda_1^3-(\lambda_1^2-1)^2\sinh(\lambda_1 kH)\right]\cosh(kH)
\right.
\\
\left.
-kL_\text{el}\lambda_1^2\left[2\lambda_1+(1+\lambda_1^2)\sinh(\lambda_1 kH)\sinh(kH)\right]\right\}\\
+k^2\lambda_1^2 L_\text{el}L_\text{cap}\left[ -2\lambda_1+2\lambda_1\cosh(kH)\cosh(\lambda_1 kH)-(1+\lambda_1^2)\sinh(kH)\sinh(\lambda_1 kH)\right]\\
+kL_\text{cap}(\lambda_1^2-1)\left[\cosh(kH)\sinh(\lambda_1 kH)-\lambda_1\sinh(kH)\cosh(\lambda_1 kH)\right] \\
+4\lambda_1(\lambda_1^2+1) -
\lambda_1(5+2\lambda^2_1+\lambda_1^4)\cosh(kH)\cosh(\lambda_1 kH)
\\+ (1+6\lambda^2_1+\lambda_1^4)\sinh(kH)\sinh(\lambda_1 kH)=0.
\label{disp}
\end{multline}
This dispersive bifurcation criterion is the main result of this
section, linking the material ($L_\text{cap}, L_\text{el}$) and geometrical
($\lambda_\text{res}$ and $K$ appearing inside the expression for
$\lambda_1$ in Eq.\eqref{l1}) parameters describing the sheared
layer, to the wavelength of the expected wrinkles (through the
non-dimensional quantity $kH = 2\pi H/\ell$, where $\ell$ is the
wrinkles' wavelength).

\begin{figure}[!ht]
\centerline{\includegraphics[height=5cm]{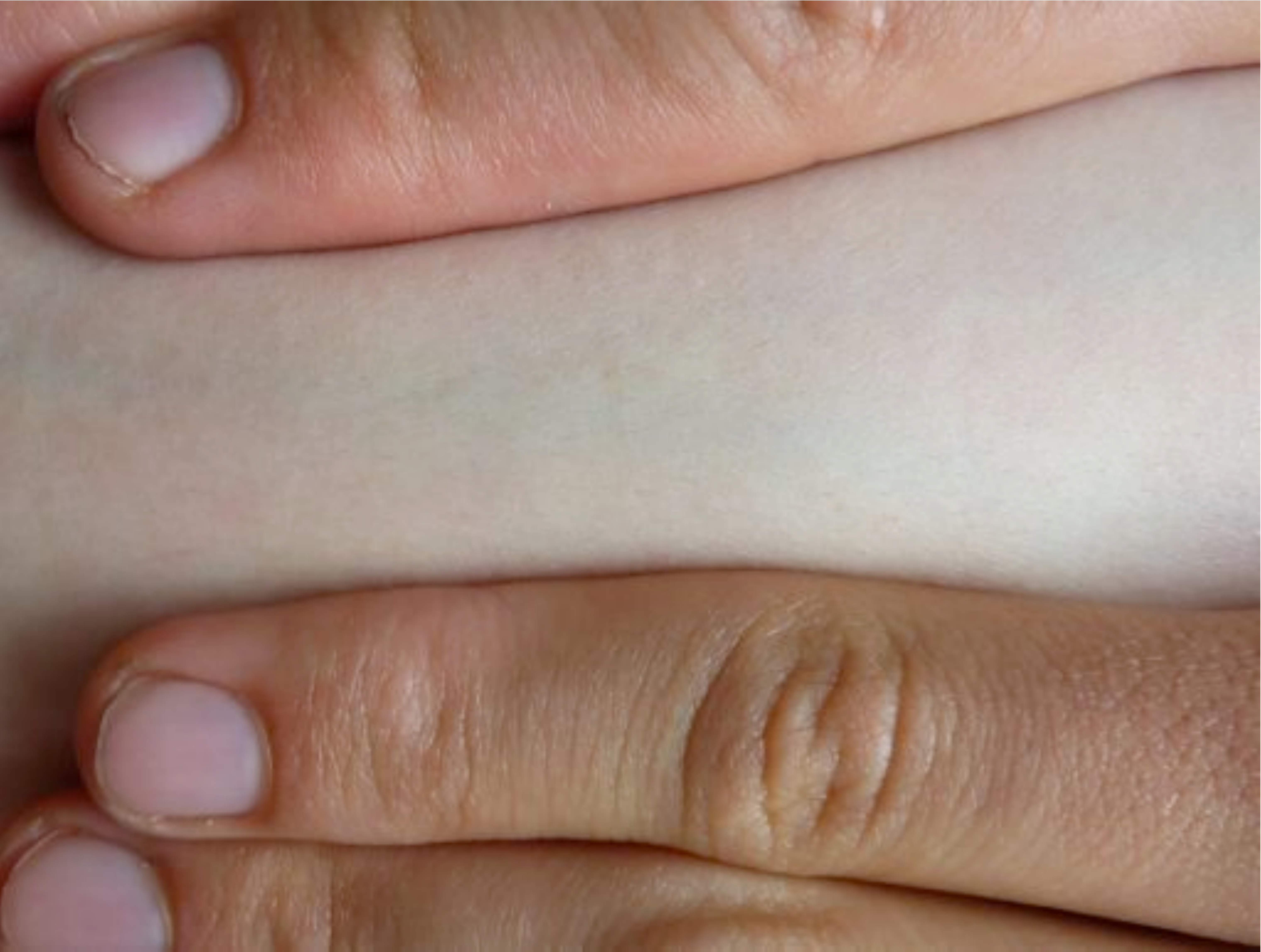} \qquad
\includegraphics[height=5cm]{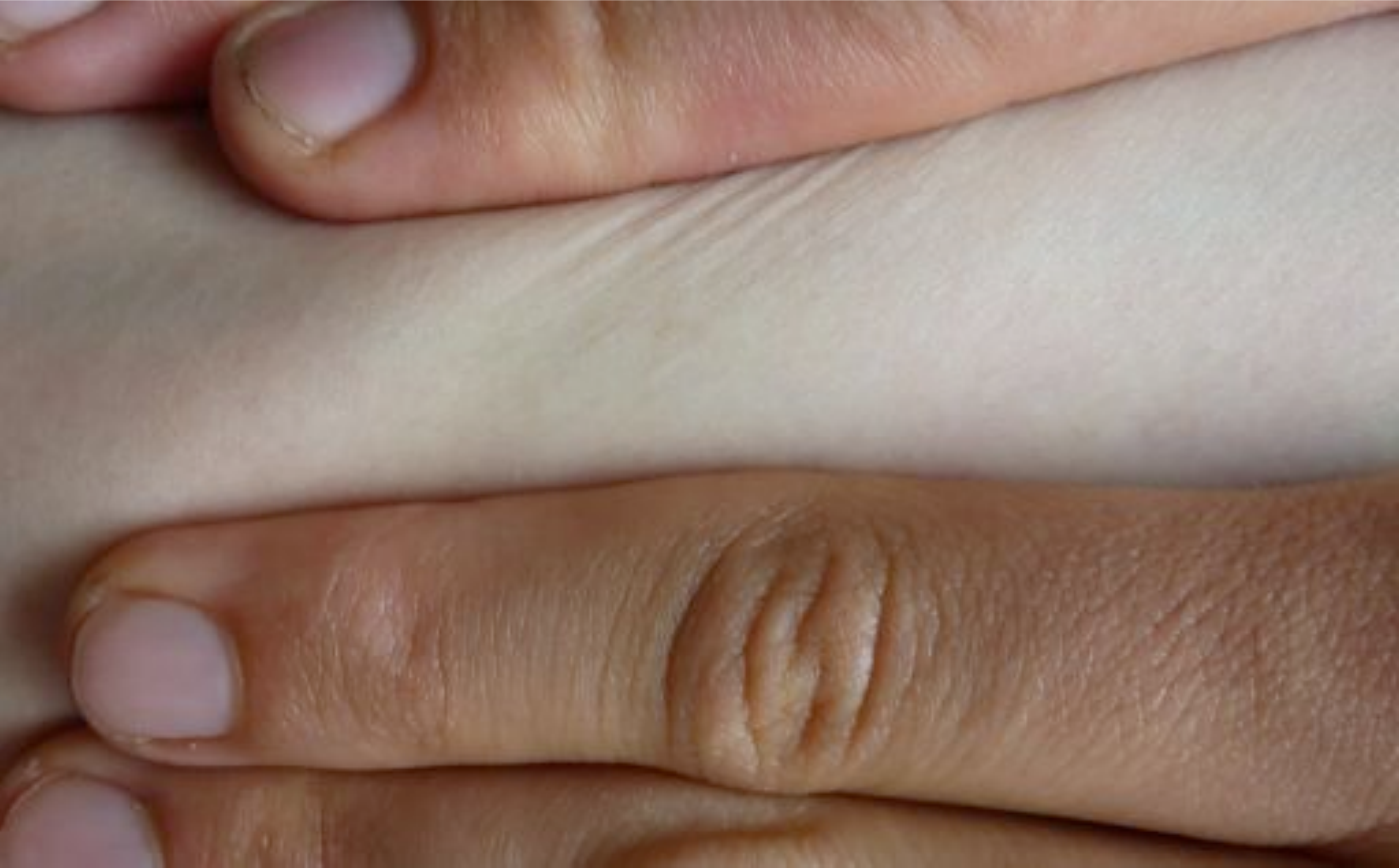}}
\caption{{\small Anecdotal evidence suggests that shearing the
forearm skin along the lines of cleavage  (which run along the
length of the arm) results in later onset of small-amplitude
wrinkles than when shearing across, compare with Figure
\ref{skin1}.}} \label{skin2}
\end{figure}

\begin{figure}[!ht]
\centerline{\includegraphics[height=5cm]{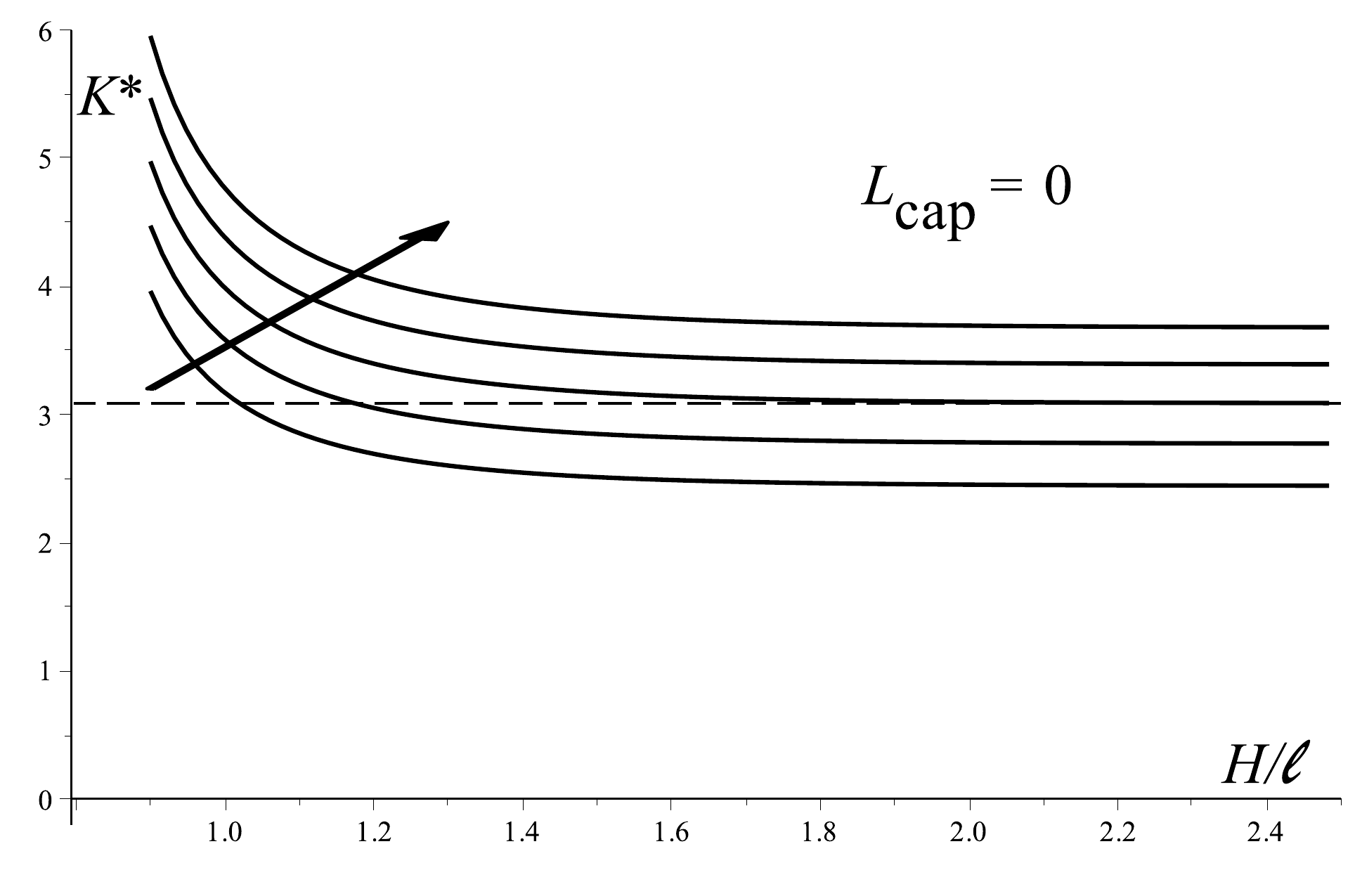} \
\includegraphics[height=5cm]{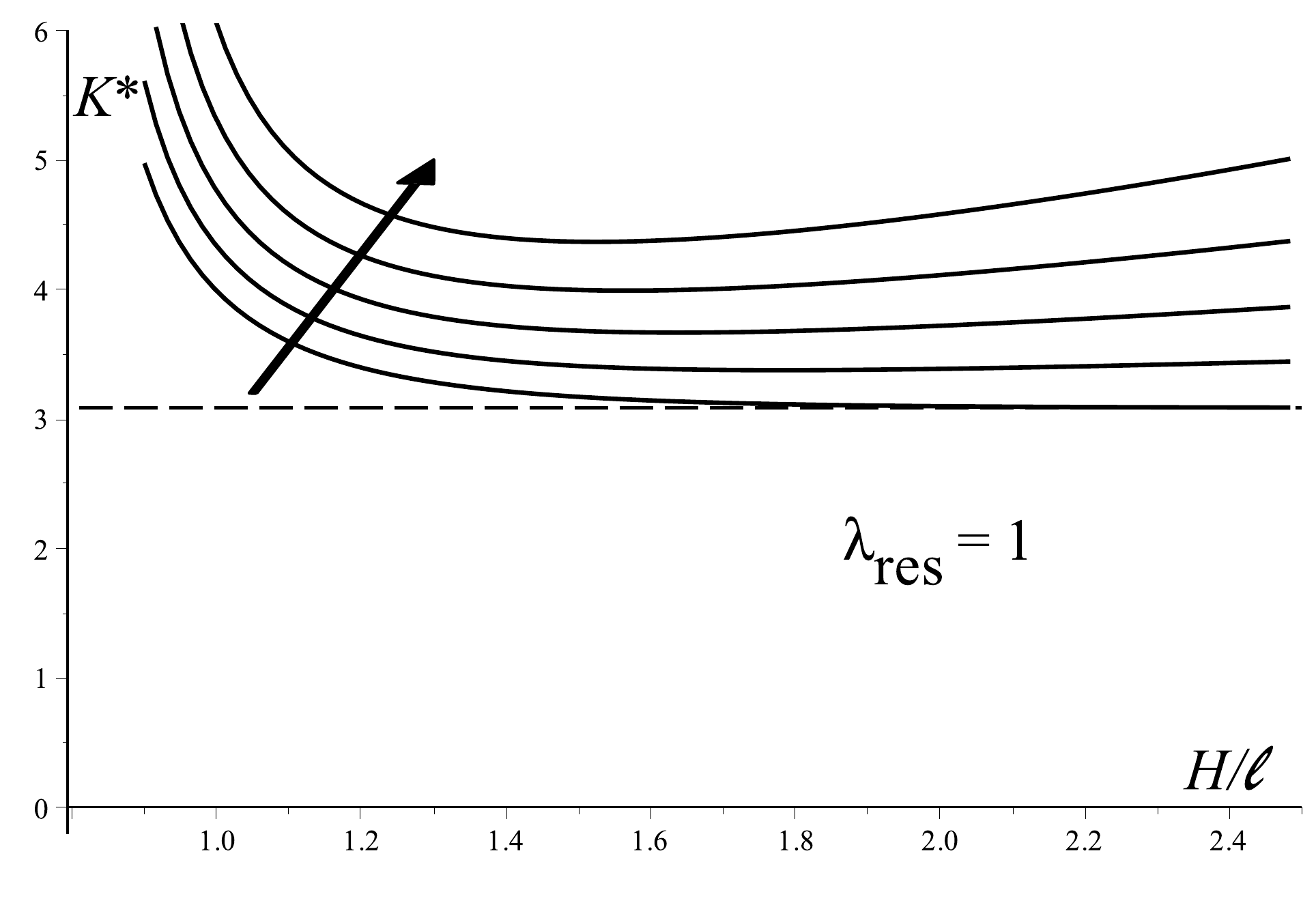}} \caption{{\small Left: In the absence of surface tension ($L_\text{cap}=0$), the layer becomes more (less) stable when sheared in the direction of tension (compression). Here $\lambda_\text{res} = 0.8,0.9,1.0,1.1,1.2$. Right: In the absence of pre-stretch ($\lambda_\text{res}=1$), the layer becomes more stable in shear when its top surface is endowed with a surface tension. Here $ L_\text{cap}/H = 0.0, 0.005, 0.01,0.015,0.02,0.025$, $L_\text{el}=0$. The horizontal line indicates the shear threshold of surface instability $K^*=3.0873$ in an unstretched half-space without surface tension.} }
\label{slab}
\end{figure}

In Figure \ref{slab} we show the separate influences of the
pre-stretch and of the surface energy. We plot the critical amount
of shear $K^*$, at which wrinkles occur, against $H/\ell$, the ratio
of the layer's thickness to the wrinkles' wavelength. We find that
as $H/\ell$ becomes small, $K^*$ increases rapidly, showing that the
layer is more and more stable: that is because for a thin slab, the
clamped boundary condition at the bottom takes precedence and
prevents the apparition of wrinkles. As soon as the layer's
thickness becomes comparable to the wrinkles' wavelength ($H/\ell >
1$), the value of the critical amount of shear tends rapidly to its
value for a semi-infinite solid (surface instability). On Figure
\ref{slab} (left), we study the influence of $\lambda_\text{res}$ in
the absence of surface energy ($L_\text{cap}=L_\text{el}=0$). We see that $K^*$
is increased when $\lambda_\text{res}>1$ and decreased when
$\lambda_\text{res}<1$. In other words, shearing along the direction
of tension requires a greater amount of shear than shearing along
the direction of compression, consistent with experimental
observations, see Figure \ref{skin2}. On Figure \ref{slab} (right),
we investigate the influence of surface tension ($L_\text{cap} >0, L_\text{el}=0$)
in the absence of pre-stretch ($\lambda_\text{res} = 1$). We see
that as $L_\text{cap}$ increases, $K^*$ also increases, indicating
that surface tension makes the layer more stable.
{A similar type of behaviour is obtained setting $L_\text{cap} =0, L_\text{el}>0$,
but it is not shown here for the sake of brevity.}
This observation is consistent with the observation that young skin does not wrinkle as
early as older skin when sheared, because it is tauter.

\begin{figure}[!ht]
\centerline{\includegraphics[height=5cm]{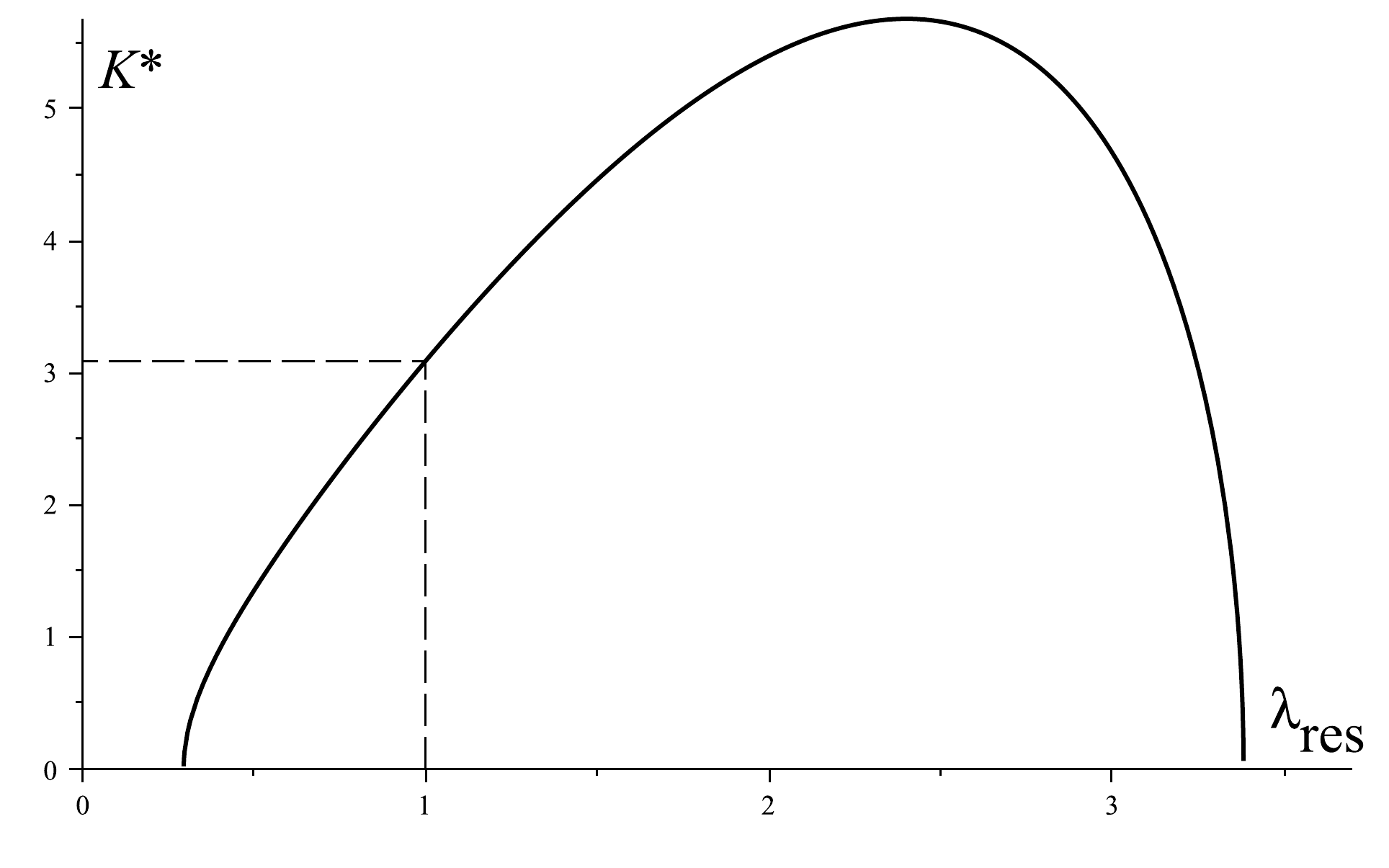} \
\includegraphics[height=5cm]{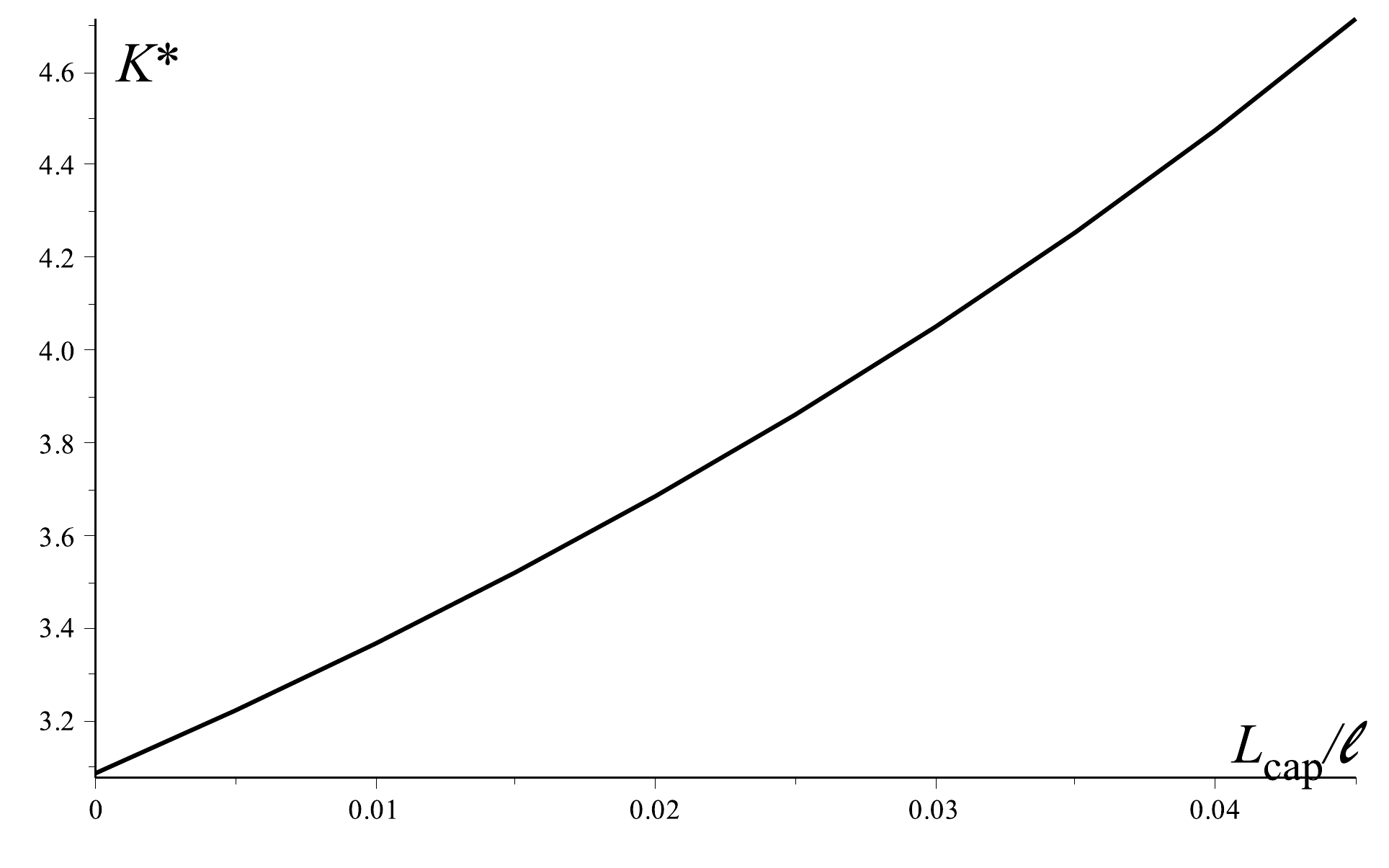}}
\caption{{\small Stability of a semi-infinite neo-Hookean subject to
a pre-stretch followed by a simple shear. Left: In the absence of
surface tension  $L_\text{cap}>0$, the stability under shear is
enhanced by a tensile pre-stretch ($\lambda_\text{res}>1$ and
decreased by a compressive pre-stretch ($\lambda_\text{res}<1$). The
dotted lines correspond to the shear threshold with no pre-stretch
($\lambda_\text{res}=1$, $K^*=3.0873$). Right: The presence of
surface tension ($L_\text{cap}>0$, $L_\text{el}=0$) allows the half-space to be sheared further before the instability criterion is met ($K^* > 3.08$, the shear threshold in the absence of surface tension.)}}
\label{half}
\end{figure}

Owing to the rapid settling of the dispersion curves to their
half-space (Biot) instability limit, we now take $kH \gg 1$ in
Eq.\eqref{disp}, while keeping $kL_\text{cap} = 2\pi
L_\text{cap}/\ell$, $kL_\text{el} = 2\pi
L_\text{el}/\ell$  finite. It then reduces to
\begin{multline}
2\pi(L_\text{el}/\ell) \lambda_1^2\left\{\lambda_1\left[1+2\lambda_1+\lambda_1^2+2\lambda_1^2\pi(L_\text{el}/\ell)\right] -\frac{\lambda_1-1}{\lambda_1+1}2\pi(L_\text{cap}/\ell)\right\}\\
+2\pi(L_\text{cap}/\ell)(\lambda_1 + 1) +(\lambda^3_1 + \lambda_1^2
+ 3\lambda_1-1)=0. \label{surf}
\end{multline}
To check for consistency, we make the link with known results. For
instance, when we neglect the surface energy in Eq.(\ref{surf}) by
taking $L_\text{el}=L_\text{cap}=0$, we recover the surface instability
criterion of plane strain \cite{destrade08},
\begin{equation}
\begin{array}{ll}
\lambda_2=\lambda_1^{-1}
=\frac{3(13+3\sqrt{33})^{1/3}}{2^{1/3}(13+3\sqrt{33})^{2/3}-2^{8/3}-(13+3\sqrt{33})^{1/3}}\simeq
3.3830.
\end{array}
\label{th}
\end{equation}
The corresponding \emph{shear threshold} $K^*$ is found by using
Eq.\eqref{xi}, as
\begin{equation} \label{half-K}
K^*=
\lambda_\text{res}\sqrt{\tfrac{2^{11/3}3^{2/3}}{3(9+\sqrt{33})^{1/3}}+\tfrac{4(45+6\sqrt{33})^{1/3}}{3}-(\lambda_\text{res}-\lambda^{-1}_\text{res})^2}
 \simeq  \lambda_\text{res}\sqrt{(3.0873)^2
-(\lambda_\text{res}-\lambda^{-1}_\text{res})^2}.
\end{equation}
Here the value  $K^*=3.0873$, obtained in the absence of a
pre-stretch ($\lambda_\text{res} = 1$), corresponds to the  shear
threshold  of surface instability for a neo-Hookean half-space as
found both theoretically by Destrade et al.\cite{destrade08} and
experimentally by Mora et al.\cite{Mora11}.
Figure \ref{half} confirms the trends found from the exact
dispersion equation. Hence the left figure (based on
Eq.\eqref{half-K}) shows that the shear threshold $K^*$ is enhanced
(half-space is more stable) by the presence of a tensile pre-stretch
and vice-versa for a compressive pre-stretch (in tension, $K^*$
eventually reaches a maximum of 5.68 at $\lambda_\text{rs} \simeq
2.40$, but this is way beyond the elastic limit of skin). Similarly,
the effect of surface tension is to increase the stretchability of
the half-space before it becomes unstable, as shown by the right
figure  (based on Eq.\eqref{surf}).

The effect of the surface energy is to fix the wavelength at
threshold, as already discussed for the surface instability of
compressed skin tissues \cite{ciarletta-MRC}. If $L_\text{cap}/H\ll
1$ and $L_\text{el}=0$, in particular, a logarithmic correction can be calculated by
series development of Eq.(\ref{disp}), reading:
\begin{equation}
 K=K^*+{\beta(\lambda_\text{res})} \frac{L_\text{cap}}{H}\log
\left[\alpha(\lambda_\text{res}) \frac{H}{L_\text{cap}}\right],
\qquad kH= \tfrac{1}{2} \log \left[\alpha(\lambda_\text{res})
\frac{H}{L_\text{cap}}\right],
\end{equation}
with
$\alpha(\lambda_\text{res})=\lambda_\text{res}(-0.0289+0.0025(\lambda_\text{res}+\lambda_\text{res}^{-2})$,
${\beta(\lambda_\text{res})}=
0.388 \lambda_\text{res} \sqrt{9.351-(\lambda_\text{res}+\lambda_\text{res}^{-2})^2}$.\\

To summarise the results of this section, we have found that the presence of a surface energy
always stabilizes the free surface and fixes the morphology of the
sheared surface.


\section{Shear instability of a fibre-reinforced skin tissue}
\label{Shear instability of a fibre-reinforced skin tissue}


The dermis of human skin is characterized by a structural
arrangement of elastin and collagen type I fibres in the
extracellular matrix, leading to an anisotropic stiffening of the
tissue. The aim of this section is to investigate how such a
material anisotropy affects the stability properties of the sheared
skin.

Let us consider a single family of fibre reinforcement oriented, in
the reference configuration, along the unit vector ${\bf M}= [\cos
\alpha, \sin \alpha, 0]^T$, defining the structural tensor ${\bf
\widehat M}={\bf M}\otimes{\bf M}$ so that $\lambda_\alpha:= ({\bf
C}:{\bf \widehat M})^{\tfrac{1}{2}}$ represents the fibre stretch, {where ${\bf
C}$ is the right Cauchy-Green deformation tensor}.
In order to build a strain measure for the fibres, we introduce the
structural invariant $I_{\alpha}$, defined as follows:
\begin{equation}
I_{\alpha}=  \left[\rm {\bf C} + {\bf C}^{-1} - 2\rm {\bf
I}\right]:{\bf \widehat M}=(\lambda_\alpha-\lambda_\alpha^{-1})^2
\end{equation}
As discussed in \cite{ciarlettaJMBBM}, this choice provides a
physically consistent deformation measure when
$\lambda_\alpha\rightarrow +\infty$ \emph{and} when
$\lambda_\alpha\rightarrow 0$, thereby allowing to account both for
compression and extension of the fibres. Accordingly, the strain
energy density of the skin tissue is defined as:
\begin{equation}
w=  \frac{\mu}{2}(I_1-3)+ {\beta} I_\alpha, \label{enaniso}
\end{equation}
where $\beta>0$ is the anisotropic elastic modulus for the fibre
reinforcement. The constitutive relation Eq.(\ref{enaniso}) ensures
strong-ellipticity of the tissue in planar deformations, a
characteristic which is not met for example for the so-called
standard model of fibre reinforcement chosen by 
\cite{destrade08}. It is a simple exercise to show that for a small
tensile strain along the direction of the fibres, we have
$\lambda_1=\lambda_\alpha=1+\epsilon$, where $|\epsilon| \ll 1$ for
the tensile stretch and $\lambda_2=\lambda_3=1-\epsilon/2$ for the
lateral stretches; then, the resulting infinitesimal stress is
$\sigma_1 = 4\mu \epsilon + 8 \beta \epsilon$, showing that (at
least in the linear regime) the ratio $2\beta/\mu$ is a measure of
the stiffness of the fibres compared to the stiffness of the matrix.

We set $\lambda_\text{res}=1$ in this section for the sake of
simplicity, so that the half-space is subject to \emph{simple shear}
only,
\begin{equation}
x=  X + K Y, \qquad y= Y, \qquad z=Z, \label{simple}
\end{equation}
with deformation gradient and principal stretches
\begin{equation}
\label{Fsimple} \mathbf F =  \begin{bmatrix} 1 & K & 0 \\ 0 & 1 & 0
\\ 0 & 0 & 1 \end{bmatrix}, \qquad \lambda_{1,2} = \pm \dfrac{K}{2}
+ \sqrt{1+\dfrac{K^2}{4}}, \qquad \lambda_3=1,
  \end{equation}
respectively. It is then easy to show that $I_\alpha=K^2$, so that
the total strain energy does not depend on fibre orientation, only
on the amount of shear $K$. The corresponding Cauchy stress tensor
does depend on fibre orientation, as follows
\begin{equation}
\boldsymbol \sigma= \mu ({\bf b}-{\bf I})+ 2{\beta} ({\bf F}{\bf M}
\otimes{\bf F}{\bf M} -{\bf F}^{-T}{\bf M} \otimes{\bf F}^{-T}{\bf
M}).
\end{equation}

Let us look for a perturbed surface wave in the form of
Eq.(\ref{sep}); to do so we need the components of the instantaneous
moduli tensor $L$ in Eq.(\ref{S1}). For the anisotropic strain
energy density $w$ defined in Eq.(\ref{enaniso}) and $\mathbf F$
given in Eq.(\ref{Fsimple}) we find the components
\begin{multline} \label{Ls}
L_{j i k l} =  \mu  \delta _{jk} b_{il} \, + \\ 2 \beta  \left( M_p
M_q  \delta _{jk} F_{lp} F_{iq} +M_p M_q \delta _{jl} F^{-1}_{pk}
F^{-1}_{q i} +M_p M_q \delta _{il} F^{-1}_{pk} F^{-1}_{q j}+M_p M_q
\delta _{ik} F^{-1}_{pj}
   F^{-1}_{q l} \right),
\end{multline}
in the coordinate system aligned with the directions of simple shear
$x$, $y$, $z$ (see Appendix for explicit expressions). Then we take
the incremental quantities to be of the form
\begin{equation}
\{ u_j, \dot S_{3j}, \dot p \}= \{U_j(kz), ik \, S_{3j}(k z), ik\,
P(kz)\}  e^{ik(\cos\theta \, x + \sin \theta \, y)}, \label{Sji}
\end{equation}
where the amplitudes are functions of $kz$ only. By a well
established procedure we can use Eqs.(\ref{inc1}-\ref{divS}) to
eliminate $P$ and write the incremental equations as a first-order
differential system known as the \emph{Stroh formulation},
\begin{equation}
{\boldsymbol \eta}' = i {\bf N}{\boldsymbol \eta} =
 i \begin{bmatrix} {\bf N}_1 & {\bf N}_2 \\ {\bf N}_3 & {\bf N}_1^T
 \end{bmatrix}
{\boldsymbol \eta}, \qquad \text{   with }  \ {\boldsymbol \eta}:=
[U_1, U_2, U_3, S_{31}, S_{32}, S_{33}]^{\mathop{T}}, \label{stroh}
\end{equation}
where the prime denotes differentiation with respect to the function
argument $kz$. Here it turns out that the blocks $\mathbf N_1$,  $\mathbf
N_2$ and $\mathbf N_3$ are symmetric (explicit expressions are given
in the Appendix).

All is in place now for a complete resolution of the surface
instability problem. There exist many strategies for this
resolution, see a partial list and references in Destrade et al.\cite{destrade08}.
Here we adopted a straightforward approach, because it turned to be
tractable numerically. Noticing that the Stroh matrix $\mathbf N$ is
constant, a solution of the system Eq.(\ref{stroh}) has the form
\begin{equation}
{\boldsymbol \eta}= {\boldsymbol \eta}_0 e^{ikqz}\qquad \text{ with
} \quad {\boldsymbol \eta}_0:= [{\bf U}_0, {\bf S}_0]^T,
\label{sol1}
\end{equation}
where ${\boldsymbol \eta}_0$ is a constant vector and $q$ are the
eigenvalues of ${\bf N}$. In order for the wrinkles' amplitude to
decay with depth, we retain the three roots $q_1$,  $q_2$, $q_3$
with positive imaginary part (i.e. $\text{Im}(q)>0$). This gives the
following general solution,
\begin{equation}
{\boldsymbol \eta}(kz)= c_1 {\boldsymbol \eta}_1 e^{ikq_1z} + c_2
{\boldsymbol \eta}_2 e^{ikq_2z} + c_3 {\boldsymbol \eta}_3
e^{ikq_3z} =
\begin{bmatrix}{\bf A}\\{\bf B}\end{bmatrix}
\begin{bmatrix}c_1\\c_2\\c_3\end{bmatrix},
\label{sol2}
\end{equation}
where $c_1$, $c_2$, $c_3$ are constants, and ${\bf A}$, ${\bf B}$
are square ($3 \times 3$) matrices built from the eigenvectors
${\boldsymbol \eta}_i$ $(i=1, 2, 3)$,  taken proportional to any
column vector of the matrix adjoint to $({\bf N}- q_i{\bf I})$. Now,
the traction-free boundary condition at $z=0$ can be written as :
\begin{equation}
{\bf S}_0=
\begin{bmatrix}S_{31}(0)\\S_{32}(0)\\S_{33}(0)\end{bmatrix} =
 {\mathbf Z} \begin{bmatrix}U_1(0)\\U_2(0)\\U_3(0)\end{bmatrix} =
{\bf Z}{\bf U}_0={\bf 0}, \label{imp}
\end{equation}
where ${\bf Z}:= - i{\bf B}{\bf A}^{-1}$ is the surface impedance
matrix. The condition for the onset of a surface instability is thus
\begin{equation}
 \det{\bf Z}=0. \label{detZ}
\end{equation}
As we do not know \emph{a priori} in which directions the wrinkles
are to appear for a given angle $\alpha$ of the fibres with respect
to the direction of shear, we need to span the entire plane and find
the angle $\theta^*$ for which the corresponding amount of shear
$K^*$ is minimal, indicating the earliest onset of wrinkling. This
is the main difference of instability behaviour between an isotropic
material (such as the material in the previous section), where the
wrinkles appear \emph{aligned} with a principal direction of
pre-deformation, and an anisotropic material, where the wrinkles may
be \emph{oblique} with respect to the direction of least stretch.

\begin{figure}[!ht]
a) \centerline{\includegraphics[height=5.5cm]{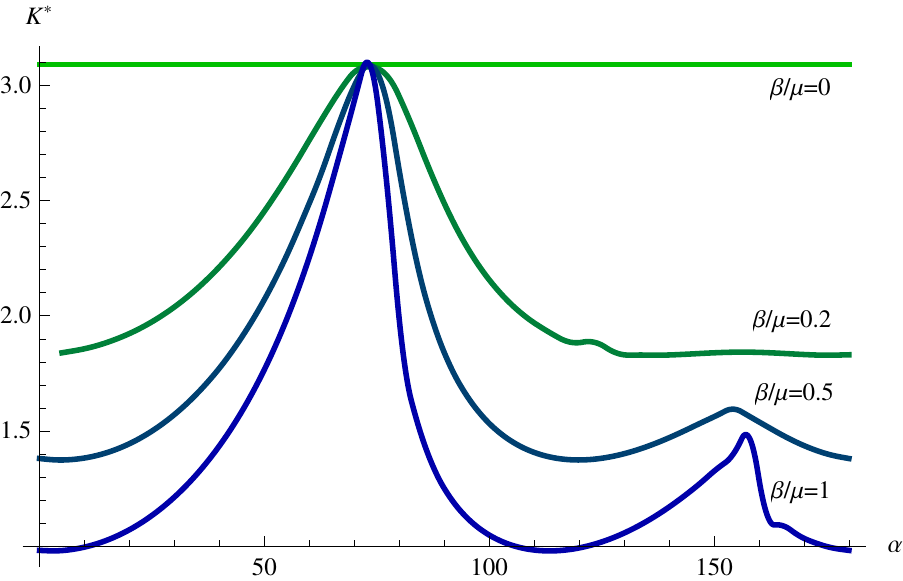} b)
\includegraphics[height=5.5cm]{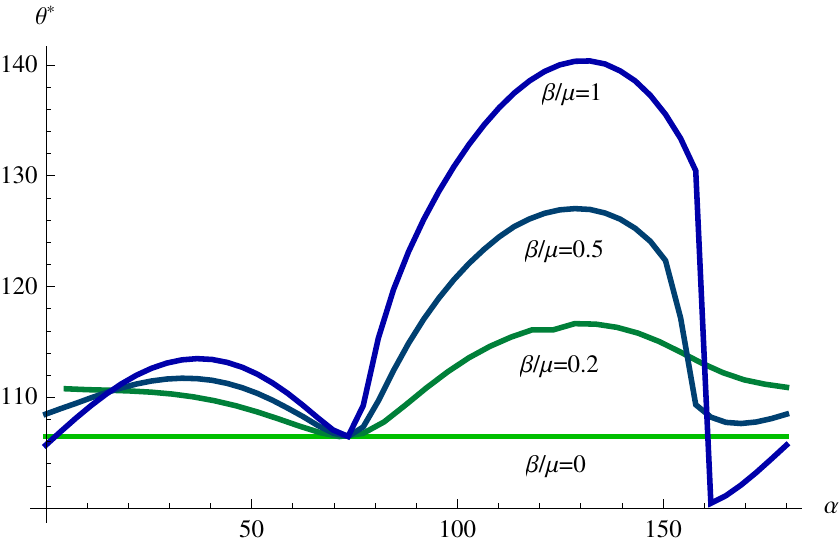}}
\caption{ (a) The critical shear strain $K^*$ as a function of the
reference fibre angle $\alpha$: The presence of fibres clearly leads
to earlier surface instability in shear. (b) The critical
instability angle $\theta^*$ as a function of the reference fibre
angle $\alpha$: These results are harder to interpret because
$\theta^*$ is defined in the current configuration and $\alpha$ in
the reference configuration. A remapping of the variations of $K^*$
and $\theta^*$ with the current fibre angle is shown in Figure
\ref{fig:Criticalalpha*}.} \label{fig:Criticalalpha}
\end{figure}

\begin{figure}[!ht]
\centerline{\includegraphics[height=7.5cm]{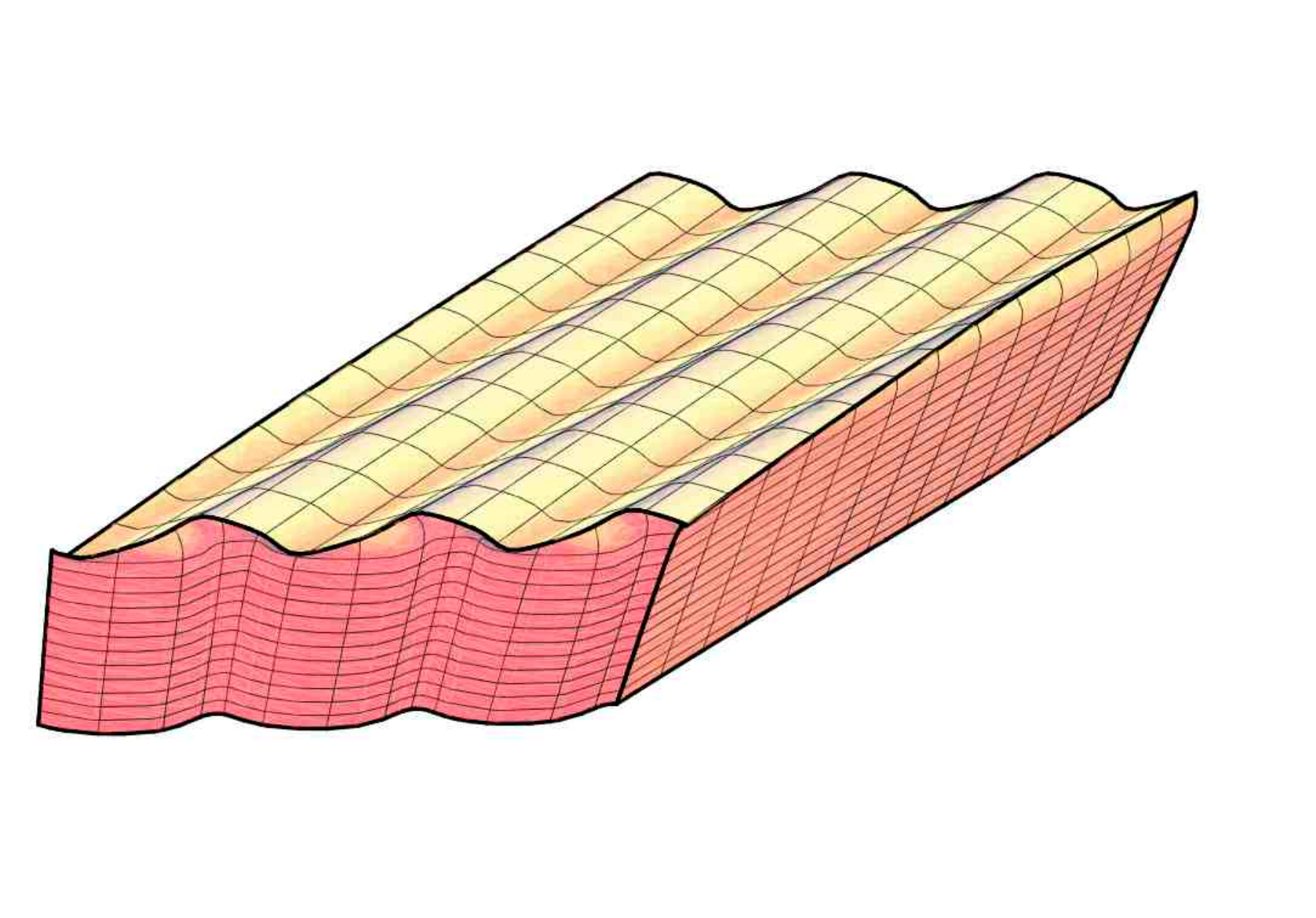}}
\caption{ When $2\beta/\mu=2.0$ (fibres are twice stiffer than
matrix in linear regime) and $\alpha = 84.5^\circ$ (fibres are
almost at right angle to the direction of shear), the first wrinkles
appear when the amount of shear reaches $K^* = 1.51$, and the
corresponding angle of the wrinkles with respect to the direction of
shear is $\theta^* = 115.2^\circ$. Note the decay of the wrinkles'
amplitude with depth.} \label{fig:Buckling3D}
\end{figure}

For our simulations, we chose material constants such that $2\beta/
\mu = 0$ (matrix alone), $0.4$ (matrix stiffer than fibres), $1.0$
(matrix as stiff as fibres), and $2.0$ (matrix softer than fibres).
For each choice of $2\beta/ \mu$ we found $K^*$ and $\theta^*$ as
functions of $\alpha$. Varying the angle $\alpha$ can be interpreted
as either varying orientation of the fibres for a shear that occurs
along a fixed axis, such as the $y$-axis in Eq.~(\ref{simple}), or
as varying the axis along which shear is taking place for a fixed
fibre direction $\alpha$.
For illustrative purposes, a typical surface buckling solution is
depicted in Figure~\ref{fig:Buckling3D}, where we chose $\beta/\mu
=1$ (fibres are twice stiffer than matrix in linear regime) and
$\alpha = 84.5^\circ$ (fibres are originally almost at right angle
to the direction of shear): there, according to
Figure~\ref{fig:Criticalalpha}, we  have $K^* = 1.51$ and $\theta^*
= 115.2^\circ$.

\begin{figure}[!ht]
\centerline{
a)
\includegraphics[height=3.5cm]{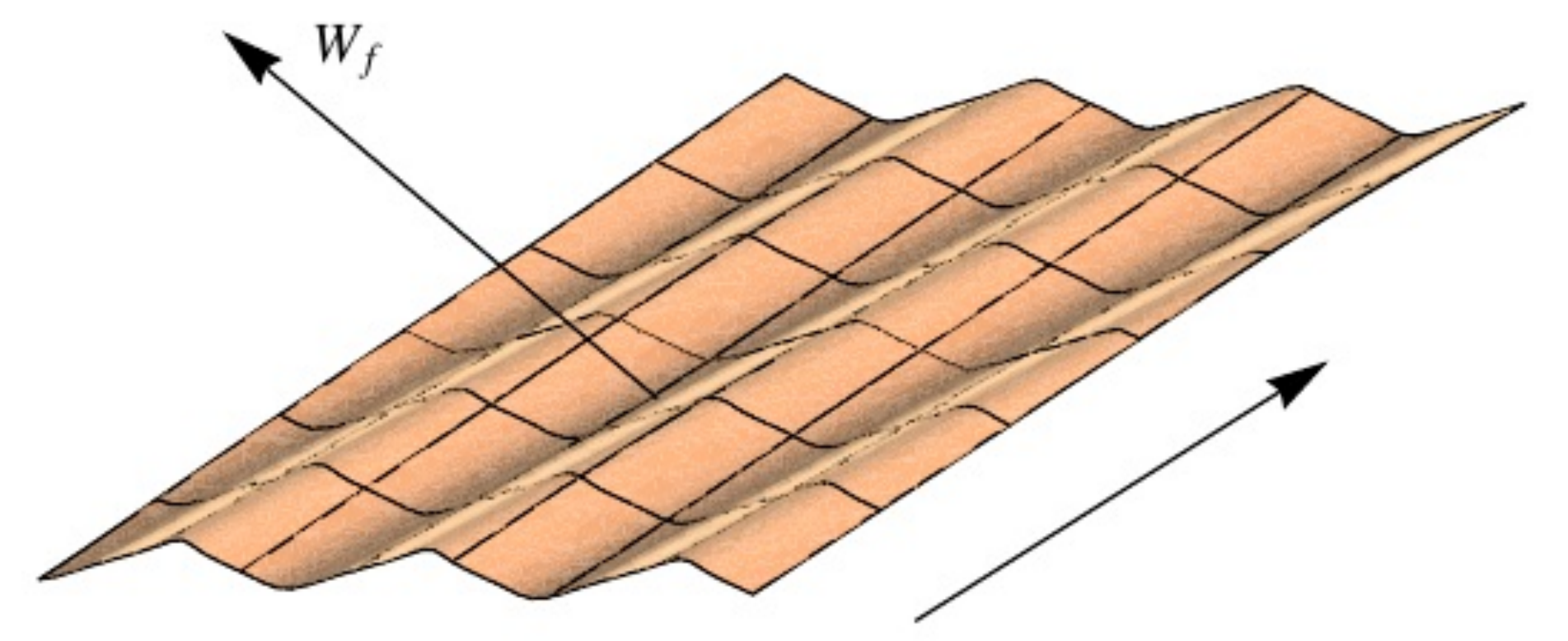}
 b)
\includegraphics[height=3.6cm]{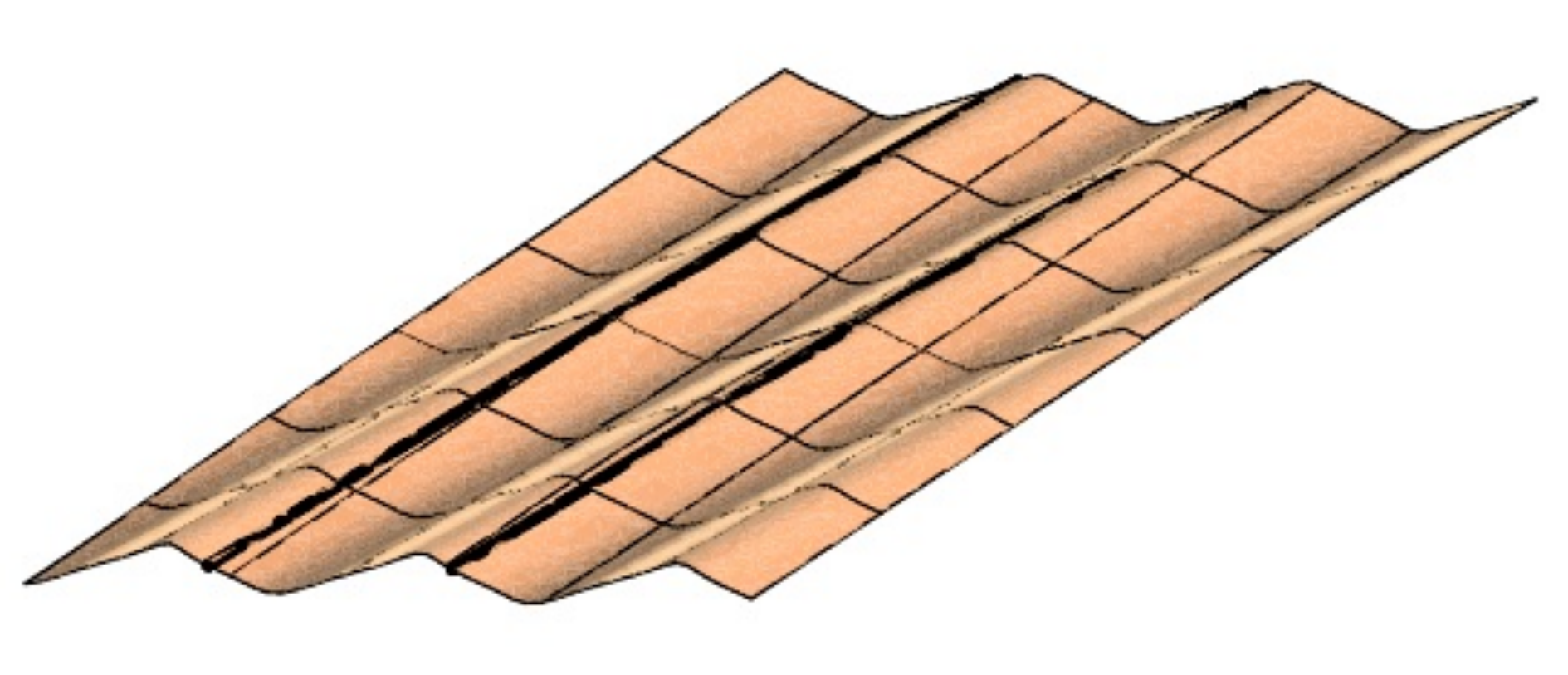}}
\caption{ a) The wavefront of the wrinkle is indicated by the vector
$\mathbf W_f$. Along the wavefront the material is alternatively elongated and compressed, orthogonal to it the material is neither elongated
nor compressed. b) The fibres are shown by bold black lines.}
\label{fig:WrinkleFront}
\end{figure}

To set the stage for analyzing the results in
Figure~\ref{fig:Criticalalpha}, we note that the first  wrinkles to
appear will occupy the least energy configuration possible  {while satisfying the zero traction boundary condition}. When the
half-space is sheared,  line elements are compressed in certain
directions and elongated in others. 
The effect of superposing a small-amplitude wrinkle is to
 { alternatively
elongate and compress} the material along the direction of the wrinkle front, i.e. in the direction $\theta$ given by Eq.~(\ref{Sji}),
 see Figure~\ref{fig:WrinkleFront}(a). 
 {This alternating behaviour, along with the zero traction boundary condition, makes it difficult to informally comprehend the influence of the wrinkle orientation, however we do notice a pattern.}
In the neo-Hookean isotropic case ($\beta=0$), the wavefront is
along the direction of greatest compression: hence here,  {wrinkling
the material in a direction under compression, due to the shear, allows the material to
release some potential energy}. In the anisotropic case  ($\beta \ne
0$),  {the presence of fibres makes the wavefront of the first wrinkle tend towards being orthogonal to the fibres.}
For instance, Figure~\ref{fig:WrinkleFront}(b) depicts the
fibre orientation for the solution in the previous figures.

\begin{figure}[!ht]
a) \centerline{\includegraphics{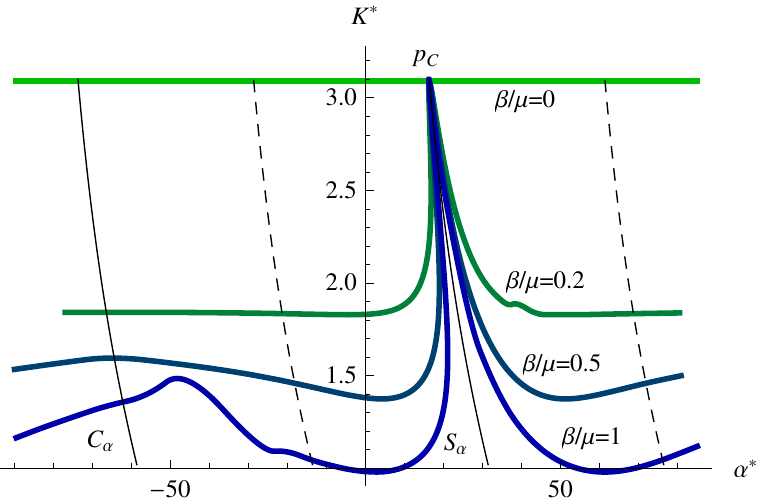} b)
\includegraphics{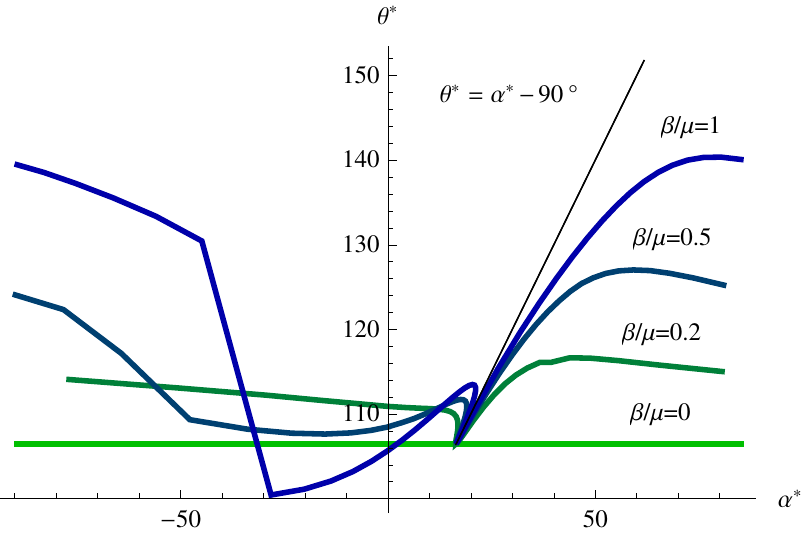}} \caption{ (a) The critical shear strain $K^*$ as a function of the current angle the fibres $\alpha^*$ with respect to the direction of shear.
(b) The critical instability angle $\theta^*$ as a function of
$\alpha^*$. The point $p_C$ indicates a surface instability state
common to all materials (independent of the material parameters).}
\label{fig:Criticalalpha*}
\end{figure}

Clearly the \emph{current} direction of fibres (in the deformed
state of finite simple shear), is closely linked to the value to the
wavefront orientation $\theta^*$. To study this relationship we
re-examine the data in Figure~\ref{fig:Criticalalpha} by mapping
$\alpha$ to the fibre orientation in the deformed body $\alpha^*$,
that is  the angle between the spatial vector ${\bf FM}$ of the current fibre
orientation and the $x$-axis, from which $\theta^*$ is also
measured. The results of this remapping are shown in
Figure~\ref{fig:Criticalalpha*}.

We first turn our attention to the plots of $K^*$ against
$\alpha^*$, see  Figure~\ref{fig:Criticalalpha*}(a). On the dashed
lines, the fibres are neither compressed or stretched. The (almost
straight) continuous black lines $S_\alpha$ and $C_\alpha$ indicate
when the fibres are aligned with the directions of greatest stretch
and greatest compression, respectively. They are given  by the
equations
\begin{equation}
  \alpha^* = \tan^{-1}\left(\lambda_2 \right), \qquad
\alpha^* = \tan^{-1} \left(\lambda_1\right),
\end{equation}
respectively, where the $\lambda$s are given in Eq.\eqref{Fsimple}
and evaluated at $K=K^*$. The $S_\alpha$ curve helps us elucidate
why there exists a point (denoted $p_C$) where all anisotropic
materials become unstable in shear at the same threshold shear
$K^*\simeq 3.0873$ as in an isotropic neo-Hookean material (where
$\beta=0$): clearly, this phenomenon occurs when the shear is such
that the fibres are aligned with the direction of greatest
stretch. Then, it turns out that great simplifications occur in the
Stroh formulation of the instability problem, and that the buckling
criterion coincides with that of the neo-Hookean model, see proof in
the Appendix. This is an artifact of our specific choice of strain
energy density in Eq.\eqref{enaniso}.


In  Figure~\ref{fig:Criticalalpha*}(b), displaying the  plots of
$\theta^*$ against $\alpha^*$, we drew the line $\theta^*=\alpha^*
-90^\circ$. Clearly, in a region close to $p_C$, the wavefront is
almost aligned with the fibres,  as is the case in an isotropic
neo-Hookean material.  As $\beta/\mu$ increases, the neighborhood of
this alignment widens, indicating that the stiffer the fibres are,
the closer the instability curves in Figure
\ref{fig:Criticalalpha*}(b) will be to the line $\theta^*=\alpha^*
-90^\circ$  {and the less the wrinkles will alter the extended fibres.}

The overall general conclusion is that stiffer fibres lead to
earlier onset of instability (notwithstanding the punctual fixing
of all curves at point $p_C$, due to {the very special case where fibres end up being aligned with the direction of greatest stretch in the deformed configuration.}) 
This result is in agreement with the
casual observation that old skin (presumably with stiffer collagen
bundles) wrinkles earlier than young skin when pinched.




\section{Discussion and Conclusion}


In this work we have investigated the occurrence of shear
instability in skin tissue within the framework of nonlinear elastic
theories.

In Section \ref{Section2}, we have considered the skin tissue as a
neo-Hookean layer of finite thickness, whilst the epidermis is
modeled \emph{as a hemitropic film with given surface energy}. Moreover, we have taken into account
the presence of the cleavage lines of skin as preferred direction of
residual stretches inside the tissue. Under these assumptions, a
linear stability analysis has been performed using the method of
incremental elastic deformations, and an analytical form of the
dispersion relation has been reported in Eq.~(\ref{disp}). The
results demonstrate that the presence of surface energy makes the
layer more stable, in the sense that it needs to be sheared more for
wrinkles to develop than when surface energy is absent
(Figure~\ref{slab}(a)). Furthermore, the surface energy fixes the
surface instability wavelength at threshold at a finite value, as
depicted in Figure~\ref{slab}(b). We have also found that wrinkles
appear earlier when the shear takes place perpendicular to the
direction of pre-stretch than when it occurs along that direction,
as confirmed by the anectodal evidences shown in Figures~\ref{skin1}
and ~\ref{skin2}.

In Section~\ref{Shear instability of a fibre-reinforced skin
tissue}, we have investigated the effect of fibre reinforcement in
the dermis layer on the shear instability characteristics. For this
purpose, we have used the polyconvex strain energy function in
Eq.(\ref{enaniso}) for modeling the transverse isotropic
reinforcement along a preferential fibre direction. A Stroh
formulation of the incremental elastic equations has been derived in
Eq.~(\ref{stroh}), and solved numerically using an iterative
technique. As shown in Figures~\ref{fig:Criticalalpha} and
\ref{fig:Criticalalpha*}, we have found that the presence of fibres
always lowers the shear threshold at which geometrical instability
happens: the stiffer the fibres, the earlier the wrinkles appear in
shear. Considering that anisotropic stiffness of skin greatly
increase with ageing \cite{agache80}, our results are in agreement
with the fact that older skin wrinkles earlier when pinched.

{
The presence of a universal point of instability at shear threshold $K^* \simeq 3.09$ when the fibres are aligned with the direction of greatest stretch $\lambda_2$, irrespective of the value of $\beta/\mu$,  can be observed on Figures \ref{fig:Criticalalpha} and \ref{fig:Criticalalpha*}. 
This anchor point is present for $\alpha_\text{cr} = \tan^{-1}(\lambda_2) \simeq
73.53^\circ$ no matter how stiff the fibres are compared to the matrix (In the Appendix we identify its origin.)
However, it represents a very special case of shear, and when we move away from the region of influence of this point, we notice that all bifurcation curves indicate a significant lowering of the shear threshold of instability (as soon as the fibres become at least as stiff as the matrix, $\beta/\mu \le 1$).
In experimental tests (see e.g. N\`i Annaidh et al.\cite{niannaidh12}), collagen fibres in human skin are determined to be at least 500 times stiffer than the elastin matrix. We may thus deduce that our model, away from the anchor point, predicts that surface instability will occur early, at low levels of shear, in line with the visual observations of Figures \ref{skin1} and \ref{skin2}.
}

In conclusion, this mathematical study of wrinkle formation in
sheared skin confirms that pinching experiments in dermatology are
useful tools to evaluate the local mechanical properties of the
tissue.




\section*{Appendix}


For an incompressible anisotropic material with strain energy
density $w$ given in Eq.\eqref{enaniso}, there are  31 non-zero
instantaneous moduli in the coordinate system aligned with the
directions of simple shear $x$, $y$, $z$ in Eq.\eqref{simple}. They
are found from Eq.\eqref{Ls} as follows.
\begin{align}
&
L_{1111} = \mu(1+K^2) + 2\beta (4\cos^2\alpha + 2K\cos\alpha\sin\alpha + K^2\sin^2\alpha), \notag \\
&
 L_{1112} = L_{1211}  = L_{2122} = L_{2221} = 4\beta \cos\alpha(\sin\alpha - K\cos\alpha),  \notag \\
&
L_{1121}= L_{1222} = L_{2111} = L_{2212} = \mu K + 2\beta (\sin 2\alpha - K\cos2\alpha),  \notag \\
&
L_{1212}=  (\mu + 2\beta)(1+ K^2),   \notag \\
&
L_{1221}  = L_{2112}= 2\beta (1-2K\cos \alpha \sin\alpha + K^2\cos^2\alpha),  \notag\\
& L_{1313} = \mu(1+K^2) + 2\beta (\cos\alpha + K\sin\alpha)^2,\notag \\
&
L_{1323} = L_{2313} = \mu K + 2\beta \sin\alpha (\cos\alpha + K\sin\alpha)^2,  \notag\\
&
L_{1331} = L_{3113} =  2\beta \cos^2\alpha, \notag \\
&
L_{1332} = L_{2331}   = L_{3123}= L_{3132}  = L_{3213}  = L_{3231} = 2\beta \cos\alpha(\sin\alpha - K\cos\alpha),   \notag \\
&
L_{2121} = \mu + 2\beta,   \notag \\
&
L_{2222} = \mu + 2\beta (4\sin^2\alpha  - 6K\cos\alpha\sin\alpha + 3K^2\cos^2\alpha), \notag \\
&
L_{2323} = \mu +  2\beta \sin^2\alpha, \notag \\
&
L_{2332}  = L_{3223}= 2\beta (\sin\alpha - K\cos\alpha)^2,   \notag\\
&
L_{3131} = \mu +  2\beta \cos^2\alpha, \notag \\
&
L_{3232}= \mu + 2\beta (\sin\alpha - K\cos\alpha)^2,   \notag \\
& L_{3333} = \mu. \notag
\end{align}
The symmetric blocks $\mathbf N_1$, $\mathbf N_2$, and $\mathbf N_3$
of the corresponding Stroh matrix $\mathbf N$ are given by
\[
-\mathbf N_1 = \begin{bmatrix}
0 & 0 & \cos \theta\\
0 & 0 & \sin \theta\\
\cos\theta & \sin\theta & 0
\end{bmatrix},
\quad \mathbf N_2= \dfrac{1}{\Delta}
\begin{bmatrix}
  L_{3232} & - L_{1332} & 0 \\
 -L_{1332} & L_{3131} & 0 \\
  0 & 0 & 0
\end{bmatrix},
\quad
-\mathbf N_3 = \begin{bmatrix} \eta & \kappa & 0\\
 \kappa & \nu & 0\\ 0 & 0 & \chi
\end{bmatrix},
\]
respectively, with
\begin{align}
&  \Delta = L_{3232}L_{3131} - L^2_{1332}
  = \mu\left[\mu + 2\beta(1-2K \cos\alpha\sin\alpha + K^2\cos^2\alpha)\right], \notag \\
& \eta=  (3\mu + L_{1111})\cos^2\theta + 2L_{1121}\cos\theta\sin\theta + L_{2121} \sin^2\theta, \notag \\
& \kappa= L_{1112} + (3\mu + L_{1221})\cos\theta\sin\theta,\notag \\
& \nu=   L_{1212}\cos^2\theta + 2L_{1121}\cos\theta\sin\theta +
(3\mu + L_{2222})\sin^2\theta,
\notag \\
& \chi= \left[\mu \sin 2\alpha +  2 \beta (\sin 2 \alpha + \sin
2\theta)\right] K  + \left[\mu \cos^2\theta + 2 \beta (\cos^2\theta
- \cos^2 \alpha)\right] K^2. \notag
\end{align}

Tremendous simplifications occur when $L$ and $\mathbf N$ are
expressed in the coordinate system aligned with the Lagrangian
principal axes and the fibres are aligned with the direction of
greatest stretch $\lambda_2$. Then, for wrinkles aligned with that
direction, we find that the Stroh matrix reads
\[
\mathbf N = \begin{bmatrix} 0 & 0 & 0 & \dfrac{\lambda_2^2}{\mu \lambda_2^2 + 2\beta} & 0 & 0 \\
0 & 0  & -1 & 0 &\dfrac{1}{\mu} & 0 \\
0 & -1 & 0 & 0 & 0 & 0 \\
-\dfrac{\mu + 2\beta}{\lambda_2^2} & 0 & 0 & 0 & 0 & 0 \\
0 & -\dfrac{\mu(1+3\lambda_2^2)}{\lambda_2^2} & 0 & 0 & 0 & -1 \\
0 & 0 & \dfrac{\mu(\lambda_2^2-1)}{\lambda_2^2} & 0 & -1 & 0
\end{bmatrix}.
\]
It clearly shows that the incremental deformation in the plane of
shear is uncoupled from the out-of-plane component. Further, the
in-plane components do not involve $\beta$ and are identical to the
components of the Stroh matrix for an isotropic neo-Hookean
material. It follows that these wrinkles appear at the critical
amount of shear $K_\text{cr} \simeq 3.0873$ found by
Destrade et al.\cite{destrade08}, independently of the value of $\beta$. The
corresponding value for the largest stretch is $\lambda_2 =
K_\text{cr}/2 + \sqrt{1 + K_\text{cr}^2/4} \simeq 3.3830$ and the
angle of the fibres in the reference configuration is
$\alpha_\text{cr} = \tan^{-1}(\lambda_2) \simeq 1.283$ rad $=
73.53^\circ$.

\end{document}